\documentclass[aps,prd,showpacs,nofootinbib,floats,floatfix,preprintnumbers,groupedaddress,twocolumn]{revtex4-2}
\usepackage{graphicx,epsfig}
\usepackage{dcolumn}
\usepackage{bm}
\usepackage{latexsym}
\usepackage{color}
\usepackage{tabularx}
\usepackage{ulem}
\usepackage{hyperref}
\usepackage{float}
\usepackage{tabularx}
\usepackage{color}
\usepackage{comment}
\usepackage{physics}

\usepackage{tikz}
\usepackage{hyperref}
\usepackage{colortbl}
\usepackage{listings}
\usepackage{amsmath,amsfonts,amssymb}
\usepackage{fancyhdr}
\usepackage{hyperref}
\usepackage{natbib}
\hypersetup{
	colorlinks   = true, 
	urlcolor     = blue, 
	linkcolor    = blue, 
	citecolor    = red 
}

\hypersetup{
	colorlinks   = true, 
	urlcolor     = blue, 
	linkcolor    = blue, 
	citecolor   = red 
}

\begin{document}
\title{Thermal nature of a generic null surface}
\author{Surojit Dalui$^a$}
\email{suroj176121013@iitg.ac.in}
\author{Bibhas Ranjan Majhi$^a$}
\email{bibhas.majhi@iitg.ac.in}
\author{T. Padmanabhan$^b$}
\thanks{Prof. T. Padmanabhan (Paddy) passed away on $17^{th}$ September, 2021. The whole idea and calculations were throughly discussed among all the authors before his decease, except  the final drafting of the manuscript.} 
\affiliation{$^a$Department of Physics, Indian Institute of Technology Guwahati, Guwahati 781039, Assam, India.\\
$^b$IUCAA, Pune University Campus, Ganeshkhind, Pune - 411 007, India.}

\date{\today}

\begin{abstract}
Dynamical properties of a generic null surface are known to have a thermodynamic interpretation. Such an interpretation is completely based on an analogy between the usual law of thermodynamics and structure of gravitational field equation on the surface. Here we materialise this analogy and show that assigning a temperature on the null surface for a local observer is indeed physically relevant. We find that for a local frame, chosen as outgoing massless chargeless particle (or field mode), perceives a ``{\it local unstable Hamiltonian}'' very near to the surface. Due to this it has finite quantum probability to escape through acausal null path which is given by Maxwell-Boltzmann like distribution, thereby providing a temperature on the surface. 
\end{abstract}

\maketitle

{\section{\label{Intro}Introduction and motivation}}
The intimate relationship between gravitational dynamics of the black hole horizon and classical thermodynamics uncovered the fact that black holes possess thermodynamic attributes like entropy \cite{Bekenstein:1973ur,Bekenstein:1974ax} and temperature \cite{Hawking:1974rv,Hawking:1975vcx,Davies:1976ei}. 
Hawking had shown \cite{Hawking:1974rv,Hawking:1975vcx} that the radiating photons are thermal in nature, and the temperature for the corresponding radiating particles was predicted as $T=\hbar\kappa/2\pi$, where $\kappa$ is the surface gravity of the black hole. 
Another phenomenon, parallel to Hawking effect, has been predicted theoretically, known as Unruh effect \cite{Unruh:1976db}. Such effect can be observed in any local arbitrary gravitational background. To make this perception clearer, we start with the {\it principle of equivalence} which allows one to construct a local inertial frame around any event in an arbitrary curved spacetime. Given the local inertial frame, one can construct a local Rindler frame, and the observer at rest in the local Rindler frame will perceive a patch of null surface as horizon with a temperature. 
This result allows us to associate thermodynamical attributions with the null surfaces, which the local Rindler observers perceive as horizons. Such a notable fact leads one to introduce observer-dependent thermodynamic variables \cite{Majhi:2012tf,Majhi:2012nq,Majhi:2017fua,Parikh:2018anm} around any event in spacetime and reinterpret the gravitational field equations near any null surface in the language of thermodynamics \cite{Padmanabhan:2009ey} or vis versa \cite{Jacobson:1995ab}. The equality between the field equations on the horizon and the thermodynamic identity has been exhibited for a wide class of models like the cases of stationary axisymmetric horizons and evolving spherically symmetric horizons in Einstein gravity \cite{Kothawala:2007em}, static spherically symmetric horizons \cite{Paranjape:2006ca}. In the Lanczos-Lovelock gravity sector, it has been studied for dynamical apparent horizons \cite{Cai:2008mh} and for generic static horizon \cite{Kothawala:2009kc}. Also, in \cite{Mohd:2013jva}, the thermodynamic identity, particularly the Clausius relation, has been established on the Local Causal Horizons using Einstein equation.


Incidentally, one can provide a thermodynamical interpretation of a gravitational field equation either by suitably projecting it on a generic null surface \cite{Parattu:2013gwa,Chakraborty:2015aja,Chakraborty:2015hna,Dey:2020tkj} or using the diffeomorphism invariance in the near null hypersurface region \cite{Chakraborty:2016dwb,Bhattacharya:2018epn,Adami:2021kvx}. For instance, Einstein's equation, contracted with null generator and corresponding auxiliary vector, yields a thermodynamical identity of the form  $\int_{S_{t}} d^{2}x T\delta_{\lambda(k)}s=\delta_{\lambda(k)}E+F\delta\lambda_{(k)}$ \cite{Parattu:2013gwa,Chakraborty:2015aja,Chakraborty:2015hna,Dey:2020tkj}, where symbols have their usual meanings and the variation can be interpreted as the change due to virtual displacement of the null surface along an auxiliary null vector $(k)$, parametrized by the parameter $\lambda$  (generalization to Lanczos-Lovelock gravity \cite{Chakraborty:2015wma} and scalar-tensor theory \cite{Dey:2021rke} has been done as well). Therefore, it has been realized that null surfaces which act as one-way membranes to a certain class of observers also possess thermodynamical attributions.               
This indispensable relation between gravity and thermodynamics led to the idea that {\it the dynamics of gravity is not fundamental in nature; rather it has emerged from the dynamics of a more fundamental theory}, just like the laws of thermodynamics of a system emerges out from the statistical dynamics of its molecules (see \cite{Padmanabhan:2009vy} for more details on the concept and aspects of emergent nature of gravity). 

However, all these thermodynamical attributions associated with the generic null hypersurface are standing on the platform of complete analogy between the laws of thermodynamics and the structure of the gravitational field equation on the surface. Therefore it is mandatory to provide a clear physical justification in order to call a geometrical quantity on the null surface as a particular thermodynamical entity. To make the concern clear, we point out that Hawking's calculation \cite{Hawking:1974rv,Hawking:1975vcx} on the emission spectrum from the black hole horizon clearly indicates the concept of the temperature of the horizon. Similarly, Unruh effect \cite{Unruh:1976db} points out the appearance of temperature on the local Rindler horizon. The lack of such robust justification in the case of generic null surface does not give us complete confidence in assigning temperature or entropy on the surface.  In this work, our particular aim is to find out the appearance of temperature without stating any analogical point of view.


In this connection, a recent development related to the thermalization mechanism of the black hole horizon is worth mentioning. It is known that the thermodynamics of a system arises from the underlying statistical theory of microstates. However, in the case of spacetime, which microscopic degrees of freedom is responsible for the emergence of temperature into the system in the presence of the horizon is not known properly. Therefore, the hunt for finding out a concrete explanation for such a feature is still under progress. The general consciousness is - to illuminate the underlying microstructure, it is mandatory to understand the physical mechanism for thermalization of the horizon.
In pursuit of achieving to find a unified reason for the origin of horizon thermodynamics, two of the authors have recently found that the thermal nature of the horizon has some connection with instability \cite{Srednicki94,Morita:2019bfr,Dalui:2019esx,Dalui:2020qpt,Majhi:2021bwo,Dalui:2021tvy} in the near-horizon region. At first glance, these two characteristics may seem to be different to each other, but their unification may become a strong candidate to answer the long-standing question about the origination of horizon thermodynamics. It has been shown through considering a model that in the presence of static spherically symmetric (SSS) black hole \cite{Dalui:2019esx,Dalui:2020qpt} or in case of a Kerr black hole \cite{Dalui:2021tvy}, an outgoing massless and chargeless particle experiences instability in the near-horizon region. In those works, it was found that a class of observers see a particle following an outgoing null path, is driven by a Hamiltonian of $xp$ kind for its near horizon motion \cite{Dalui:2019esx,Dalui:2020qpt,Dalui:2021tvy}. It may be noted that such a Hamiltonian is unstable in nature. Moreover, in the quantum scale, this unstable Hamiltonian provides the temperature into the system, which exactly matches with Hawking's expression. This noticeable fact leads to a conjecture -- ``{\it the local instability in the vicinity region of the horizon acts as the source of the temperature of the system}'' (see also \cite{Dalui:2019esx,Dalui:2020qpt}). The noticeable feature of the model is it requires mainly the information of a suitable near horizon null path of the particle, which does not crucially consider the underlying symmetry of the spacetime, like the presence of a timelike Killing vector.



Therefore, keeping the above feature in mind, we want to extend the spirit of the aforesaid model in the case of any generic null surface. We will see that it works very well in this case.
Let us now summarize the main outcomes and features of our present investigation. 
\begin{itemize}
	\item First, we start with a massless scalar field and using Klein-Gordon (KG) equation in the field-theoretic approach, we obtain that in the semiclassical limit, the system Hamiltonian for the outgoing mode in the near null hypersurface region comes out to be of $xp$ kind. 
	\item Using the conventional idea of tunnelling mechanism \cite{Srinivasan:1998ty,Parikh:1999mf,Banerjee:2008sn,Banerjee:2009wb,Majhi:2011yi} for this Hamiltonian, we obtain that quantum mechanically, the object can see the null surface as a thermal system. The importance of this observation is that it not only shows an application of tunnelling methodology for a more general background but also verifies the fact that the association of temperature on a generic null surface is very much physically acceptable.
	\item Therefore, our results shed light on the intrinsic reason why temperature is associated with the null surface. The local unstable feature can be responsible for such thermalization.
	\item{Finally, in the present work, unlike the previous ones (SSS and Kerr), we deal with a metric where any intrinsic symmetry, like time translational invariance of spacetime, is absent. Therefore it shows a wide applicability and generality of our developed approach.}
\end{itemize}

The paper is organised as follows:  In Section \ref{GNC metric and Null hypersurfaces}, we first introduce the metric adapted to Gaussian null coordinates (GNC) in the neighbourhood region of any generic null hypersurface and some precursory properties of the line element. Next, in Section \ref{KG eqn} we construct the Hamiltonian of the real scalar field mode using the Klein-Gordon (KG) equation in the near null hypersurface region. In Section \ref{Transeverse average} we calculate the transverse coordinate average of the Hamiltonian of the system for the implementation of tunneling approach. In the next part in Section \ref{Tunneling and Thermality}, we apply the tunneling formalism for our system. In the final Section \ref{Conclusion} we discuss the key features of our work.



{\section{\label{GNC metric and Null hypersurfaces} Null Hypersurfaces in GNC coordinates}}
This paper intends to investigate the possible thermodynamics properties of an arbitrary null surface. Therefore, our first objective is to describe the neighbourhood region of that null-hypersurface. A preferable choice of coordinate system exists to narrate in this context, known as Gaussian null coordinates (GNC), in analogy with Gaussian normal coordinates. Usually, the Gaussian normal coordinates are constructed by extending the coordinates on a non-null hypersurface to a spacetime neighbourhood using geodesics normal to the surface. However, this construction does not apply to a null surface because the normal geodesics lie itself on it. Therefore, an uniquely defined auxiliary null geodesics are introduced with some certain conditions in order to construct the Gaussian null coordinates, which we shall discuss in the latter part of this section. An elaborate description and detailed discussion on the construction of this coordinate system and how metric is constructed in this coordinate system can be found in \cite{Moncrief:1983xua,Hollands:2006rj,Morales:2008,Parattu:2015gga}. Here, in order to keep the clarity of our work, we shall briefly describe some precursory construction of this coordinate system in a more intuitive way.  

The primary objective is to construct a coordinate system around any null surface in any spacetime. Consider a smooth null surface $N$ in a four-dimensional spacetime manifold $M$, where $g_{ab}$ represents the metric on $M$. We start with a spacelike 2-surface $\zeta$, on $N$ and the coordinates on $\zeta$ are introduced as $(x^{A})$ for $A=1,2$, i.e. $(x^{1},x^{2})$. Now, in order to construct the null surface $N$ using null geodesics, one cannot have these null geodesics lying on the spacelike surface as any vector lies on the spacelike surface has to be spacelike in nature. Therefore, the next coordinate parameter should be introduced in such a way that one must move away from this spacelike surface along any of these null geodesics. Here, we introduce that particular parameter $v$, not necessarily affine, along the null geodesics increasing in the future direction with the condition $v=0$ on $\zeta$. Therefore, any point on $N$ in the neighbourhood of $\zeta$ can be labelled by the coordinates $(v,x^{1},x^{2})$, where $(x^{1},x^{2})$ corresponds to the label given to the null geodesic passing through that point and $v$ is the chosen parameter at that point. The ``neighbourhood'' corresponds to that sufficiently small region where the geodesics do not cross each other or do not form any caustics. Let us call the future directed vector field tangent to the null geodesics, $\partial/\partial v$, as $\mathbf{l}$. Now, our spacetime is four-dimensional, so it is time for us to introduce the fourth coordinate. With the help of a new set of null geodesics, we can construct the coordinate chart in the surroundings of the null-hypersurface. Introducing a unique null vector $k^{a}$ which is situated at each point on the null surface and satisfying these conditions (i) $k^{a}k_{a}=0$, as it is a null vector; (ii) $l^{a}k_{a}=-1$ which suggests that $k^{a}$ sticks out from each point of the  null surface instead of lying on the surface; and (iii) $X^{a}_{A}k_{a}=0$ where $X_{A}=\partial/\partial x^{A}$ are the basis vectors correspond to the coordinates $(x^{1},x^{2})$. The choice of the third condition shows that our auxiliary null vector is uniquely defined. As this vector $k^{a}$ points out off the null surface, it can be used to go off the null surface. The null geodesics emitted from each point on the null surface in the direction of $k^{a}$ are labelled by the coordinates $(v,x^{1},x^{2})$, of that point. We choose a parameter $r$ along this null geodesic and $r=0$ (we choose again) represents the null surface and $\mathbf{k}=-\partial/\partial r$. Therefore, the chosen affine parameter $r$ can be assigned in the coordinate chart $(v,r,x^{1},x^{2})$ in the neighbourhood of the null hypersurface means up to the regions where geodesics do not reach a caustic.

After introducing this coordinate system, we shall introduce the metric adapted to this coordinate system in the neighbourhood region of any arbitrary null hypersurface. The construction of this metric has been detailed in \cite{Hollands:2006rj,Morales:2008}. However, we shall recall here some of the essential properties of this metric. The line element adapted in this context takes the following form
\begin{eqnarray}
ds^{2}=-2r\alpha dv^{2}+ 2dv dr - 2r\beta_{A}dv dx^{A}+\mu_{AB}dx^{A}dx^{B}~.
\label{GNC metric}
\end{eqnarray}  
Now, one can see from (\ref{GNC metric}) that there are two null surfaces which are at $r=0$ and at $v=$ constant. Additionally, $r=$ constant where the constant is non-zero represents the timelike surfaces. Besides, the metric components $\alpha,~\beta_{A}$ and $\mu_{AB}$ are the smooth functions of all the coordinates and $\mu_{AB}$ is the transverse $(D-2)$-dimensional Riemannian metric on the spacelike surface $\zeta$. We have here a set of null vectors as we mentioned earlier $l^{a}=(l^{v},l^{r},l^{x^{A}})=(1,0,\mathbf{0})$ and the unique auxiliary one, i.e. $k^{a}=(0,-1,\mathbf{0})$. Among these two we can think of $l^{a}$ as the future-outgoing null vector and $k^{a}$ as the future-ingoing null vector. The covariant components of these two vectors are $l_{a}=(-2r\alpha,1,-r\beta_{A})$ and $k_{a}=(-1,0,\mathbf{0})$. Therefore, the normal $l_{a}=\partial_{a}r$ to the $r=0$  surface will be a null vector. Moreover this generates the $r=0$ null surface around which we will do our all analysis.

In order to strengthen the motivation for choosing our desired null hypersurface at $r=0$, let us first write the static spherically symmetric (SSS) metric in analogous to the GNC form \cite{Chakraborty:2015aja}
\begin{eqnarray}
ds^{2}=-f(r)dv^{2}+2dvdr+\mu_{AB}dx^{A}dx^{B}~.\label{SSS metric}
\end{eqnarray}
One can see $\beta_{A}=0$ in this case comparing it with Eq. (\ref{GNC metric}). Now, in the near horizon region we have $f(r)\simeq 2\kappa(r-r_{H})$ where $\kappa=f'(r_{H})/2$ where the prime denotes the derivative with respective to $r$. Therefore, comparing (\ref{SSS metric}) with Eq. (\ref{GNC metric}) we obtain $\alpha\equiv\kappa$ and the position of the horizon $r=r_{H}$ is equivalent to the position of the null surface at $r=0$ in case of GNC metric (\ref{GNC metric}). Therefore, in this context the null surface at $r=0$ is reminiscent to the black hole horizon. In our earlier works \cite{Dalui:2019esx,Dalui:2020qpt} we studied the particle dynamics near the horizon and obtained that particle trajectory becomes unstable near it. Following that essence our objective in this work is to extend this idea in case of any generic null hypersurface and thereby we particularly choose the null surface which is at $r=0$. The null normal $l_{a}$ which is at $r=0$ is given by $l_{a}=(0,1,\mathbf{0})$ is the generator of the null surface providing $l^{a}l_{a}=0$ at $r=0$. Therefore, for our examination in the near null hypersurface region we shall consider the limit $r\rightarrow 0$ whenever necessary.

In the present context, our prime aim is to explain how temperature is physically associated with any generic null hypersurface. In order to do so, the first objective is to find out the emission probability of a particle through the null hypersurface. So, we adopt one of the familiar techniques known as the tunneling method \cite{Srinivasan:1998ty,Parikh:1999mf,Banerjee:2008sn,Banerjee:2009wb,Majhi:2011yi}. This formalism predicts the probability of escaping the particle through the null hypersurface. Therefore, to evaluate the tunneling probability first, we need to find out the paths of the particle for both ingoing and outgoing cases. Hence, we acquire the Hamiltonian-Jacobi (HJ) formalism in this case, just like our earlier works \cite{Dalui:2019esx,Dalui:2020qpt,Dalui:2021tvy}, in order to find out the Hamiltonian in the near null hypersurface region for both ingoing and the outgoing cases.
{\section{\label{KG eqn} Hamiltonian: field description}}

Considering the massless real scalar field $\phi$, from the Klein-Gordon (KG) equation $\square\phi=0$ under the background of metric (\ref{GNC metric}) yields
\begin{eqnarray}
&&\partial_{v}\left(\sqrt{\mu}\partial_{r}\phi\right)+\partial_{r}\left(\sqrt{\mu}\partial_{v}\phi\right)+\partial_{r}\left[\sqrt{\mu}\left(2r\alpha+r^{2}\beta^{2}\right)\partial_{r}\phi\right]
\nonumber
\\
&&+\partial_{r}\left(\sqrt{\mu}~r\beta^{A}\partial_{A}\phi\right)+\partial_{A}\left(\sqrt{\mu}~r\beta^{A}\partial_{r}\phi\right)
\nonumber
\\
&&+\partial_{A}\left(\sqrt{\mu}\mu^{AB}\partial_{B}\phi\right)=0~,
\label{KG equation}
\end{eqnarray}
where $\mu$ is the determinant of the induced metric $\mu_{AB}$.
Now, we start with the standard ansatz for the scalar field of a particle as (see also \cite{Srinivasan:1998ty})
\begin{eqnarray}
\phi=\mathcal{A}(v,r,x^{A}) e^{-\frac{i}{\hbar}S(v,r,x^{A})}\label{field ansatz}~,
\end{eqnarray}
where $S(v,r,x^{A})$ is the HJ action and with respect to the HJ action we define the four-momentum as
\begin{eqnarray}
\frac{\partial S}{\partial x^{a}}=p_{a}~.\label{four momentum}
\end{eqnarray}
Now, expanding $S(v,r,x^{A})$ in the powers of $\hbar$ we find,
\begin{eqnarray}
S(v,r,x^{A})&=&S_{0}(v,r,x^{A})+\hbar S_{1}(v,r,x^{A})+ \hbar^{2} S_{2}(v,r,x^{A})\nonumber\\
&& +....\nonumber\\
&=&S_{0}(v,r,x^{A})+\sum_{i}\hbar^{i}S_{i}(v,r,x^{A})~,
\label{Paddy1}
\end{eqnarray}
where $i=1,2,3,\dots$. The terms from $\mathcal{O}(\hbar)$ onward are treated as the quantum corrections over the semi-classical value $S_{0}.$ However, our analysis is restricted only upto the semi-classical limit, i.e. $\hbar\rightarrow 0$. Therefore, the higher order terms of $\hbar$ can be neglected in the semi-classical limit. At this point, we define $-\partial S_{0}/\partial v=-p_{v}=H$, where $H$ is the (semi-classical) Hamiltonian of the system. 

Now, the main interest lies in the near $r\rightarrow 0$ region because that is where the dynamics of our massless scalar modes will be studied. The probability of crossing the mode across the null surface will be our main quantity to find out. One of the well-known techniques to study such quantity in this region is tunneling formalism. The principal way of implementing the tunneling formalism is the HJ method \cite{Srinivasan:1998ty}. To implement this idea, we need to identify the ingoing and the outgoing modes near the null hypersurface region. The outgoing mode is moving from $r<0$ region (call as ``{\it{inside}}'') to $r>0$ region (call as ``{\it{outside}}'') and vice versa for the ingoing one.
Now, applying $\hat{p}\equiv -i\hbar\partial/\partial r$ on (\ref{field ansatz}) we shall have the momentum eigenvalue as $-\partial S_{0}/\partial r$ in the semi-classical limit. For the outgoing case 
we have the positive momentum eigenvalue, which means one must have  $\partial S_{0}/\partial r<0$. Similarly, for the ingoing case, we have the negative momentum eigenvalue, i.e. $\partial S_{0}/\partial r>0$. After this identification, we need to calculate the HJ actions for both outgoing and the ingoing modes and implement these expressions to calculate the tunneling probability to cross the null hypersurface.      

In the semi-classical limit (i.e. $\hbar\rightarrow 0$), keeping only the leading order terms, we obtain the following form of the Eq. (\ref{KG equation}):
\begin{eqnarray}
2(\partial_{v}S_0)(\partial_{r}S_0)&&+(2r\alpha+r^{2}\beta^{2})(\partial_{r}S_0)^{2}+2r\beta^{A}(\partial_{A}S_0)(\partial_{r}S_0)\nonumber\\
&&+\mu^{AB}(\partial_{B}S_0)(\partial_{A}S_0)=0~.\label{KG with semi classical limit}
\end{eqnarray}
Here we see from Eq. (\ref{KG with semi classical limit}) that $\partial_{r} S_0$ has two solutions which are
\begin{eqnarray}
\partial_{r}S_0=-\frac{\partial_{v}S_0 + r\beta^{A}(\partial_{A}S_0)}{2r\alpha + r^{2}\beta^{2}}&&\pm\Bigg[\left(\frac{\partial_{v}S_0 + r\beta^{A}(\partial_{A}S_0)}{2r\alpha + r^{2}\beta^{2}}\right)^{2}\nonumber\\
&&-\frac{\mu^{AB}(\partial_{A}S_0)(\partial_{B}S_0)}{2r\alpha+r^{2}\beta^{2}}\Bigg]^{\frac{1}{2}}\label{radial momentum solution}~.
\end{eqnarray}
Among these two solutions, one corresponds to the outgoing mode, and the other one corresponds to the ingoing one. Let us get going to identify them. 

First, we need to find out the leading order solutions of $\partial_{r}S_0$ in the near null hypersurface region.
Considering the negative sign of Eq. (\ref{radial momentum solution}), we obtain the leading order term in $r\rightarrow 0$ limit as (for details please see Appendix \ref{App1})

\begin{eqnarray}
\partial_{r}S_0\Big|_{-}=-\frac{\partial_{v}S_0}{\alpha^{(0)} (v,x^{A})r}~.\label{outgoing S_r}
\end{eqnarray}  
According to our definition of the Hamiltonian of the system (which we have defined earlier below Eq. (\ref{Paddy1})) we can write Eq. (\ref{outgoing S_r}) as
\begin{eqnarray}
\partial_{r}S_0\Big|_{-}=\frac{H}{\alpha^{(0)}(v,x^{A})r}~.\label{outgoing S_0 in terms of Hamiltonian}
\end{eqnarray}
For the initial position of the mode is at  \textit{`inside'}, we have $\partial_{r}S_0<0$ when $H>0$. Therefore, as we mentioned earlier, the momentum direction is in the outward direction.  Hence, the negative sign corresponds to the outgoing mode. So, we can write the Hamiltonian for the outgoing mode in the near null hypersurface region as
\begin{eqnarray}
H=\alpha^{(0)}(v,x^{A})rp_{r_{out}}\label{outgoing Hamiltonian}
\end{eqnarray}  
where $p_{r_{out}}$ is the outgoing momentum in $r$ direction.


Similarly, considering the positive sign of Eq. (\ref{radial momentum solution}) we obtain the leading order term that survives at $r\rightarrow 0$ limit is 
\begin{eqnarray}
\partial_{r}S_0\Big|_{+}=-\frac{1}{2}\frac{\mu^{(0)AB}(\partial_{A}S_0)(\partial_{B}S_0)}{\partial_{v}S_0}~,
\label{ingoing S_r}
\end{eqnarray}
where $\mu^{(0)AB}$ is the first term of the expansion of $\mu^{AB}$ about $r=0$.
Therefore, in terms of the Hamiltonian, we can write the above equation as
\begin{eqnarray}
\partial_{r}S_0\Big|_{+}=\frac{1}{2}\frac{\mu^{(0)AB}(\partial_{A}S_0)(\partial_{B}S_0)}{H}~.\label{ingoing S_r Hamiltonian term}
\end{eqnarray}
This implies for $H>0$ we have $\partial_{r}S_0>0$. Therefore the momentum direction, in this case, is in the inward direction. So, it corresponds to the ingoing mode, and the expression of Hamiltonian in this case is 
\begin{eqnarray}
H=\frac{1}{2}\frac{\mu^{(0)AB}p_{A}p_{B}}{p_{r_{in}}}~\label{ingoing Hmiltonian}~,
\end{eqnarray}
where $p_{r_{in}}$ is the ingoing momentum in $r$ direction. 

From the expression of the outgoing Hamiltonian (\ref{outgoing Hamiltonian}) one can see that the outgoing mode suffers a singularity at $r=0$. In contrast, the ingoing mode does not experience such a thing (see Eq. (\ref{ingoing Hmiltonian})). This interesting observation has significant implications in the calculation of the tunneling probability, as we shall see in the later parts.
Furthermore, in Appendix \ref{Lagrangian formalism} we also varified the form of the outgoing Hamiltonian (Eq. (\ref{outgoing Hamiltonian})) in the particle description through the Lagrangian formalism.     



{\section{\label{Transeverse average} Transverse coordinate average of the Hamiltonian}}
Next, we want to explore the consequences of this classical Hamiltonian in the quantum tunnelling picture. It may be worth to point out here that in earlier calculations \cite{Dalui:2019esx,Dalui:2020qpt,Dalui:2021tvy} the spacetime metric was static or stationary and hence $\alpha^{(0)}$ was constant. However, this is not the case here.

However before proceeding for executing the tunneling formalism, let us prepare the stage for implementing it on our Hamiltonian. The structure of the Hamiltonian for the outgoing particle in the near null hypersurface region is multidimensional in this case due to the presence of $\alpha^{(0)}(v,x^{A})$. Therefore, we have a situation where the case of a multidimensional tunneling has appeared. Multidimensional tunnelling event has been discussed in \cite{Razavy Book} for the usual physical systems. One of the proposals to calculate tunnelling probability is to do calculation on an average potential by considering averaging over directions except one. Hence following this idea and since tunnelling occurs radially just across the null surface, we read the transverse coordinates average of the Hamiltonian: $\bar{H}=\int H\sqrt{\mu}~ d^{2}x^{A} / \int \sqrt{\mu} d^{2}x^{A}$. This yields
\begin{eqnarray}
\bar{H}=\bar{\alpha}(v)rp_{r_{out}}\equiv \bar{E}~,
\label{AverageH}
\end{eqnarray}
where $\bar{\alpha}(v)$ is defined as 
\begin{eqnarray}
\bar{\alpha}(v)=\frac{\int \alpha^{(0)}(v,x^{A})\sqrt{\mu}d^{2}x^{A}}{\int\sqrt{\mu}d^{2}x^{A}}~.\label{average alpha}
\end{eqnarray}
Here one thing is to be mentioned that during the consideration of the average of the Hamiltonian $H$, the integrating average applied only on $\alpha^{(0)}(v,x^{A})$ because both $r$ and $p_{r_{out}}$ are independent of $x^{A}$. We shall use this average Hamiltonian (\ref{AverageH}) in the next section in order to investigate the thermalization of our null surface through tunneling formalism.


{\section{\label{Tunneling and Thermality} Tunneling and Thermality}}

We start by calculating the HJ action for the outgoing object. Choosing the integration limit from $r=-\epsilon$ to $r=\epsilon$ for the outgoing object where $\epsilon>0$ and is a very small number suggests that the outgoing object crosses the null hypersurface from just \textit{`inside'} to just \textit{`outside'} in the vicinity of the null hypersurface. Therefore, we obtain the HJ action for the outgoing species (field mode or the particle) as
\begin{eqnarray}
S_{out}&=&\frac{\bar{E}}{\bar{\alpha}(v)}\int_{-\epsilon}^{\epsilon}\frac{dr}{r}+\int p_{v_{out}}dv + \int p_{A_{out}}dx^{A}~.\label{outgoing action}
\end{eqnarray}
From the above integration it can be seen that the first integration term will contain the imaginary part as there exists a singularity at $r=0$ and other two integrations will contribute in the real part of the total integration. Also in the first integration we have pulled out the term $\bar{E}/\bar{\alpha}(v)$ as it is constant of motion (please see Appendix \ref{App2}). Hence, after performing the integration in Eq. (\ref{outgoing action}) we obtain
\begin{eqnarray}
S_{out}=-\frac{i\pi\bar{E}}{\bar{\alpha}(v)}+\text{Real part}~.
\end{eqnarray}     
In a similar way, we can calculate the HJ action for the ingoing species also. However, in this case, the action does not contain any singularity at $r=0$ (see Eq. (\ref{ingoing Hmiltonian}) and Eq. (\ref{momenta Lagrangian formalism})); thus, it turns out to be
\begin{eqnarray}
S_{in}=\text{Real quantity}~.\label{S_in}
\end{eqnarray} 
Accordingly, the probability for the outgoing object crossing the null hypersurface turns out to be
\begin{eqnarray}
P_{out}&\sim& \Big|e^{-\frac{i}{\hbar}S_{out}}\Big|^2 
\nonumber
\\
&\varpropto& \exp\left(-\frac{2\pi \bar{E}}{\hbar\bar{\alpha}(v)}\right)~,
\label{Pout}
\end{eqnarray} 
whereas the probability of crossing the null hypersurface for the ingoing one is $P_{in}\sim 1$. Therefore, the tunneling probability comes out to be
\begin{eqnarray}
\Gamma(v)=\frac{P_{out}}{P_{in}}\sim \exp\left(-\frac{2\pi \bar{E}}{\hbar\bar{\alpha}(v)}\right)~.
\label{Pout/Pin}
\end{eqnarray}
This particular expression of the tunneling probability is similar to Boltzmann factor. Therefore can be considered as thermal in nature with the temperature of the system is identified as
\begin{eqnarray}
T(v)=\frac{\hbar\bar{\alpha}(v)}{2\pi}\label{temperature}~.
\end{eqnarray}
However, this very expression of the temperature is not a constant; instead, it is a function of $v$. It means at every other $v=constant$ null hypersurface near $r=0$ region, the observer will feel different values of temperature of the system for every different value of $v$. This is a reflection of the evolving nature of our null surface which corresponds to a non-equilibrium situation. We will come back to this point again in the next section.
{\section{\label{Conclusion}Discussion}}
Let us summarise the results obtained in the present work. We started this work by addressing the fact that gravitational field equations near any null surface in an arbitrary space-time reduce to a thermodynamic identity, and it generalises the results previously available in the context of the horizon. Our prime motive was to find out the underlying reason for this noticeable fact in order to convey the cause why thermodynamical attributions are associated with any arbitrary null hypersurface. We start our calculations using the KG equation in the field-theoretic approach, and in the semi-classical limit, we obtain that the system Hamiltonian for the outgoing mode in the near null hypersurface region comes out to be of $xp$ kind. In the appendix, the same has been explored in the Lagrangian formalism for a massless outgoing particle as well.
In the context of thermality, we proceed with the conventional idea of tunneling mechanism, and after implementing the tunneling formalism in the near null hypersurface region, we obtain that our system is thermal in nature. However, the system temperature we found, in this case, is not constant; rather, it is a function of the timelike coordinate, unlike the previous results of the black hole horizons (SSS BH and the Kerr one).

Now, let us discuss the key features of our work in a more detailed manner. Our results justify the fact how temperature can be associated with any generic null hypersurface. Some earlier works predicted that the emergence of thermality into the system has a close connection with the {\it local} instability of the system in the context of horizon \cite{Morita:2019bfr,Dalui:2019esx,Dalui:2020qpt,Dalui:2021tvy}. This connection previously showed that if the Hamiltonian of the system turns out to be an unstable one in the classical scale, this instability may lead to the thermality of the system in the quantum scale \cite{Dalui:2019esx,Dalui:2020qpt,Dalui:2021tvy}. Here, we came across the Hamiltonian, which consists of a probed massless and chargeless species near any generic null hypersurface and the structure of the outgoing Hamiltonian, in this case, turns out to be of $xp$ kind (see Eq. (\ref{outgoing Hamiltonian})). 

Note that such specific Hamiltonian turns out to be that of an inverted harmonic oscillator (IHO) in a new set of canonical variables $(X,P)$: $x=\frac{1}{\sqrt{2}}(P-X)$ and $p=\frac{1}{\sqrt{2}}(P+X)$ \cite{Book1} and IHO potential is inherently unstable. This implies that our present outgoing species locally feel an instability due to the presence of null surface at $r=0$. This fact can also be realised through the divergence of radial momentum $p_{r_{out}}$ at $r=0$ for a given value of $\bar{E}$ (see Eq. (\ref{AverageH})). 
Such peculiar instability provides a noticeable feature in the quantum regime. To escape through the potential ($\sim xp$) the outgoing object needs to tunnel through a complex path as it experiences a singularity exactly at $r=0$. Moreover, as we noticed in the calculation, $r=0$ singularity (which is also the key for aforesaid instability) led to our main expression of tunneling probability (\ref{Pout/Pin}).
Usually, the time-reversal invariance demands that the emission probability is equal to that for the absorption
process proceeding backwards in time and vice versa. Whereas our present result is not consistent with this. Therefore the present observation shows that the probability of emission of particles through the null surface at a certain time is different from the probability of absorption of particles by the surface at that time.
Hence it is more likely for a particular region to gain particles than lose them. 
Moreover, the exponential behaviour of our result portrays the thermal nature of the system. This thermality comes into the picture only because of this peculiar singularity at $r=0$, which originates due to the specific structure of the outgoing Hamiltonian in the null surface regime. Therefore, we feel that the local instability in the near null hypersurface region may be the reason for making the system thermal at the quantum scale.

Previously, this connection between instability and thermality was established only in specific cases containing horizons. Here we generalise the same for a generic null hypersurface. Therefore, we feel that the present discussion may unfold the deeper reason for having the thermodynamical quantities of not only horizon but also for any generic null surface at the quantum level. Moreover, this work also represents one of the important applications of tunneling mechanism for more general background.

Finally, we make a comment on the conceptual aspect of defining thermodynamics on a generic null surface which is an evolving one. Thermodynamics for an equilibrium system is well established. In contrast, our null surface can not be considered as an equilibrium one. Therefore the concept of temperature and corresponding zeroth law etc., are not consistent with equilibrium thermodynamics. Instead, we need to invoke ``non-equilibrium'' definitions of these thermodynamic quantities. This subject is not fully established, but there are a few suggestions and advancements. 
A point to be noted is that if the system is in non-equilibrium steady states, different thermometers, sensitive to different degrees of freedom (DOF), will show different temperature readings, which lead to the difficulty of defining only one temperature for these systems \cite{Casas:2003}. Hence, the equilibrium version of the zeroth law does not work in its full glory. However, a restricted validation of zeroth law can be considered here, and in that case, the temperature must be defined with respect to some specified DOF. For instance, if a system is composed of two subsystems and they have different DOF, then corresponding to each DOF one can define a temperature. Consequently, the zeroth law is valid within that particular DOF. 
This, in turn, gives rise to different ``local'' temperatures in the system as it consists of different degrees of freedom. In this local sense, the law of thermodynamics and the thermodynamics parameters can be defined, but that will be accompanied by heat flux, temperature gradient etc., among different DOF (see discussion in Section 4.1 of \cite{Casas:2003} for details). Of course, for the ``global'' equilibrium, this wipeout.
Now, comparing with that situation, we see our system also obeys the characteristics of the non-equilibrium steady-state situation where our null hypersurface is evolving with the changing value of $v$. Therefore, we expect that the temperature of our system will be defined following the same concept as it is defined in non-equilibrium situations.
A proposal for defining the effective temperature was suggessted by S. Weinberg in the case of a non-equilibrium system of photons by relating absorption rate coefficient $\Lambda$ and the stimulated emission coefficient $\Omega$ (see discussion in Section 6.2 of \cite{Casas:2003} for details):
\begin{equation}
\frac{\Omega}{\Lambda} = e^{-\frac{E}{T_{\textrm{eff}}}}~.
\end{equation}
In the present discussion, we have adopted the same spirit in order to identify the temperature of the null surface. However, the status of the zeroth law for our system is still an open question as the complete knowledge of the degrees of freedom for our system is yet to be explored and therefore needs further investigation. However, we feel that the lack of a complete theory of non-equilibrium thermodynamics at present will keep us at bay to get full justification of thermodynamics of a null surface. 
On the other hand, if we consider that the evolution of the null hypersurface is quasi-static in nature, then our temperature can be justified through equilibrium thermodynamics by considering that the surface is at equilibrium at each instant.


\appendix
\begin{widetext}
\section*{Appendices}
{\section{\label{App1}Derivation of Eq. (\ref{outgoing S_r}) and Eq. (\ref{ingoing S_r})}}
In the near null hypersurface region, i.e. $r\rightarrow 0$ limit, $\alpha(v,r,x^{A})$ can be expanded (using Taylor series expansion)
\begin{eqnarray}
\alpha(v,r,x^{A})=\alpha^{(0)}(v,x^{A})+\alpha^{(1)}(v,x^{A})r+\mathcal{O}(r^{2})~.
\end{eqnarray}  
Now, looking back to Eq. (\ref{radial momentum solution}) we can rewrite it as
\begin{eqnarray}
\partial_{r}S=&&-\frac{\partial_{v}S + r\beta^{A}(\partial_{A}S)}{2r\alpha + r^{2}\beta^{2}}\pm\left(\frac{\partial_{v}S + r\beta^{A}(\partial_{A}S)}{2r\alpha + r^{2}\beta^{2}}\right)\Bigg[1-\frac{\mu^{AB}(\partial_{A}S)(\partial_{B}S)}{\left(\partial_{v}S + r\beta^{A}(\partial_{A}S)\right)^{2}}\left(2r\alpha+r^{2}\beta^{2}\right)\Bigg]^{\frac{1}{2}}~.
\end{eqnarray}
Now, at $r\rightarrow 0$ the above equation turns into
\begin{eqnarray}
\partial_{r}S&\simeq &-\frac{\partial_{v}S + r\beta^{A}(\partial_{A}S)}{2r\alpha + r^{2}\beta^{2}}\pm\left(\frac{\partial_{v}S + r\beta^{A}(\partial_{A}S)}{2r\alpha + r^{2}\beta^{2}}\right)\Bigg[1-\frac{1}{2}\frac{\mu^{AB}(\partial_{A}S)(\partial_{B}S)}{\left(\partial_{v}S + r\beta^{A}(\partial_{A}S)\right)^{2}}\left(2r\alpha+r^{2}\beta^{2}\right)\Bigg]\nonumber\\
&\simeq&-\frac{\partial_{v}S + r\beta^{A}(\partial_{A}S)}{2r\alpha + r^{2}\beta^{2}}\pm\Bigg[\left(\frac{\partial_{v}S + r\beta^{A}(\partial_{A}S)}{2r\alpha + r^{2}\beta^{2}}\right)-\frac{1}{2}\frac{\mu^{AB}(\partial_{A}S)(\partial_{B}S)}{\left(\partial_{v}S + r\beta^{A}(\partial_{A}S)\right)}\Bigg]~.\label{A3}
\end{eqnarray}
Considering the negative sign solution of $\partial_{r}S$ we obtain from Eq. (\ref{A3})
\begin{eqnarray}
\partial_{r}S\Bigg|_{-}=-2\frac{\partial_{v}S + r\beta^{A}(\partial_{A}S)}{2r\alpha + r^{2}\beta^{2}} + \frac{1}{2}\frac{\mu^{AB}(\partial_{A}S)(\partial_{B}S)}{\left(\partial_{v}S + r\beta^{A}(\partial_{A}S)\right)}~.\label{A4}
\end{eqnarray}
The first term of Eq. (\ref{A4}), using the expansion of $\alpha(v,r,x^{A})$, in the near null hypersurface region reduces to
\begin{eqnarray}
\frac{\partial_{v}S + r\beta^{A}(\partial_{A}S)}{2r\alpha + r^{2}\beta^{2}}&\simeq & \frac{\partial_{v}S+r\beta^{A}(\partial_{A}S)}{2r\alpha^{(0)}\left(1+\frac{r\alpha^{(1)}}{\alpha^{(0)}}+\frac{r\beta^{2}}{2\alpha^{(0)}}\right)}\nonumber\\
&=& \frac{\partial_{v}S+r\beta^{A}(\partial_{A}S)}{2r\alpha^{(0)}}\left(1-\frac{r\alpha^{(1)}}{\alpha^{(0)}}-\frac{r\beta^{2}}{2\alpha^{(0)}}\right)~.
\end{eqnarray}
Similarly, from the second term of Eq. (\ref{A4}), in the near null hypersurface region we obtain
\begin{eqnarray}
\frac{\mu^{AB}(\partial_{A}S)(\partial_{B}S)}{\left(\partial_{v}S + r\beta^{A}(\partial_{A}S)\right)} &\simeq & \frac{\mu^{AB}(\partial_{A}S)(\partial_{B}S)}{\partial_{v}S }\left(1-\frac{r\beta^{A}(\partial_{A}S)}{\partial_{v}S}\right)~.
\end{eqnarray}
Hence, putting the approximated values of these two terms in Eq. (\ref{A4}) we obtain the only leading order term in the near null hypersurface region
\begin{eqnarray}
\partial_{r}S\Bigg|_{-}=-\frac{\partial_{v}S}{2r\alpha^{(0)}(v,x^{A})},
\end{eqnarray} 
i.e. Eq. (\ref{outgoing S_r}) in our main text. In the similar manner we can also obtain the expression of $\partial_{r}S|_{+}$.
Taking the positive sign solution in Eq. (\ref{A3}) and considering the leading order term at $r\rightarrow 0$ limit we end up getting Eq. (\ref{ingoing S_r}). 
{\section{\label{Lagrangian formalism} Hamiltonian: Particle description in Lagrangian formalism}}
 In Section \ref{KG eqn}, using HJ formalism we land up to a particular Hamiltonian structure (Eq. (\ref{outgoing Hamiltonian})) of the outgoing scalar mode in the near null hypersurface region. Here we like to find out whether the same Hamiltonian structure can be obtained using the Lagrangian of a particle. 

Consider the Lagrangian $L=\sqrt{-g_{ab}\dot{x}^{a}\dot{x}^{b}}$ where $\dot{x}^{a}=dx^{a}/dv$. Since we are considering a massless particle, for convenience $v$ has been chosen here as the affine parameter for the geodesics of the particle. Therefore, under the background of metric (\ref{GNC metric}) we obtain the form of the Lagrangian of the system as
\begin{eqnarray}
L=\left[ 2r\alpha-2\dot{r}+2r\beta_{A}\dot{x}^{A}-\mu_{AB}\dot{x}^{A}\dot{x}^{B}\right]^{\frac{1}{2}}
\end{eqnarray}
where the expressions of the corresponding momentum components are
\begin{eqnarray}
p_{r}&=&-\frac{1}{\left[ 2r\alpha-2\dot{r}+2r\beta_{A}\dot{x}^{A}-\mu_{AB}\dot{x}^{A}\dot{x}^{B}\right]^{\frac{1}{2}}};\\
p_{A}&=&\frac{r\beta_{A}-\mu_{AB}\dot{x}^{B}}{\left[ 2r\alpha-2\dot{r}+2r\beta_{A}\dot{x}^{A}-\mu_{AB}\dot{x}^{A}\dot{x}^{B}\right]^{\frac{1}{2}}}~.
\end{eqnarray}
Therefore, we obtain the Hamiltonian of the system as
\begin{eqnarray}
H=\frac{1}{2p_{r}}\left[(2r\alpha + r^{2}\beta^{2})p_{r}^{2} + 2r\beta^{A}p_{A}p_{r}+(1+p_{A}^{2})      \right]~.\label{Hamiltonian app}
\end{eqnarray}
The above expression of the Hamiltonian (\ref{Hamiltonian app}) reveals that there are two solutions of $p_{r}$ in trems of $H$ and $p_A$. It is evident that one solution of $p_{r}$ corresponds to the outgoing particle while the other one corresponds to the ingoing one. 
Now, in the near null hypersurface region ($r\rightarrow 0$) considering only the leading order terms in these two solutions of $p_{r}$, we obtain
\begin{eqnarray}
p_{r}\Big|_{-}=\frac{H}{\alpha^{(0)}(v,x^{A})r}~~~~\text{and}~~~~p_{r}\Big|_{+}=\frac{1}{2}\left(\frac{1+p_{A}^{2}}{H}\right)\label{momenta Lagrangian formalism}
\end{eqnarray}
where $(-)$ and $(+)$ sign represents the -ve and the +ve sign solutions of $p_{r}$ of the quadratic equation (\ref{Hamiltonian app}) respectively. Therefore, we can see that the expression of the -ve sign solution of $p_r$ exactly matches with Eq. (\ref{outgoing S_0 in terms of Hamiltonian}) which we identified as the momentum in $r$ direction for the outgoing mode, i.e. $p_{r_{out}}$. So, it is evident that the -ve sign solution of $p_{r}$ corresponds to the momentum of the outgoing particle and we obtain the similar structure of the outgoing Hamiltonian in the near null hypersurface region (see Eq. (\ref{outgoing Hamiltonian})). 



Therefore, using the particle description in Lagrangian formalism we obtain the similar expression of the outgoing Hamiltonian in the near null hypersurface region as we obtained in the field mode description in Section \ref{KG eqn}. Whereas for the ingoing particle the Hamiltonian structure in the near null hypersurface region may differ in those two descriptions (see (\ref{ingoing Hmiltonian}) and Eq.  (\ref{momenta Lagrangian formalism})) but their natures are same as the ingoing particle does not suffer any singularity at $r=0$.


{\section{\label{App2} Conserved quantity $\bar{H}/\bar{\alpha}(v)$ }}
Now, we have the near null hypersurface Hamiltonian for the outgoing particle, i.e.
	\begin{eqnarray}
	H=\alpha^{(0)}(v,x^{A})rp_{r_{out}}
	\end{eqnarray}
	and after averaging out the transverse coordinates we have
	\begin{eqnarray}
	\bar{H}=\bar{\alpha}(v)rp_{r_{out}}~.
	\end{eqnarray}
	Now, let us check the variation of $\bar{H}/\bar{\alpha}(v)$ with respect to some affine parameter $\lambda$, i.e.
	\begin{eqnarray}
	\frac{d}{d\lambda}\left(\frac{\bar{H}}{\bar{\alpha}(v)}\right)&=&\frac{d}{d\lambda}\left(rp_{r_{out}}\right)\nonumber\\
	&=&\dot{r}p_{r_{out}}+r\dot{p}_{r_{out}}\label{variation wrt lambda}
	\end{eqnarray}
	where $.\equiv \frac{d}{d\lambda}$. Now, from the Hamilton's equations of motion we obtain
	\begin{eqnarray}
	\dot{r}=\frac{\partial\bar{H}}{\partial p_{r_{out}}}=\bar{\alpha}(v)r
	\end{eqnarray}
	and the other one is
	\begin{eqnarray}
	\dot{p}_{r_{out}}=\frac{\partial\bar{H}}{\partial r}=-\bar{\alpha}(v)p_{r_{out}}~.
	\end{eqnarray}
	Now, putting the values of $\dot{r}$ and $\dot{p}_{r_{out}}$ in Eq. (\ref{variation wrt lambda}) we obtain
	\begin{eqnarray}
	\frac{d}{d\lambda}\left(\frac{\bar{H}}{\bar{\alpha}(v)}\right)&=&\bar{\alpha}(v)rp_{r_{out}}+\left(-\bar{\alpha}(v)rp_{r_{out}}\right)\nonumber\\
	&=&0~.
	\end{eqnarray}
	It tells that the quantity $\bar{H}/\bar{\alpha}(v)=\bar{E}/\bar{\alpha}$ is conserved during the motion of the particle under the average Hamiltonian $\bar{H}$.

\end{widetext}

	
\end{document}